\documentclass[usenatbib]{mn2e}
\usepackage{graphicx}
\usepackage{times}

\newcommand{\jed}[1]{~{\rm #1}}

\title[Limits on additional planets in HD~189733]{Tight constraints on the
existence of additional planets around HD~189733
\thanks{Based on observations made with the Nordic Optical Telescope,
operated on the island of La Palma jointly by Denmark, Finland, Iceland,
Norway, and Sweden, in the Spanish Observatorio del Roque de los
Muchachos of the Instituto de Astrof\'{i}sica de Canarias.}
\thanks{Based on observations made with the William Herschel
Telescope operated
on the island of La Palma by the Isaac Newton Group in the Spanish
Observatorio del Roque de los Muchachos of the Instituto de Astrof\'{i}sica de
Canarias.}}
\author[M.~Hrudkov\'{a} et al.]{M.~Hrudkov\'{a}$^{1,2,3}$\thanks{E-mail:
marie@tls-tautenburg.de}, I.~Skillen$^{2}$, C.~R.~Benn$^{2}$,
N.~P.~Gibson$^{4,5}$, D.~Pollacco$^{4}$, D.~Nesvorn\'{y}$^{6}$, 
\newauthor
T.~Augusteijn$^{7}$, S.~M.~Tulloch$^{2}$ and Y.~C.~Joshi$^{4,8}$\\  
$^{1}$Astronomical Institute of the Charles University,
V Hole\v{s}ovi\v{c}k\'{a}ch~2, CZ-180~00~Praha 8, Czech Republic\\
$^{2}$Isaac Newton Group of Telescopes,
Apartado de Correos 321, E-387~00 Santa Cruz de la Palma, Canary Islands,
Spain\\
$^{3}$Th\"{u}ringer Landessternwarte Tautenburg, Sternwarte 5, D-07778
Tautenburg, Germany\\
$^{4}$Queen's University Belfast, University Road, BT7 1NN, Belfast, UK\\
$^{5}$School of Physics, University of Exeter, EX4 4QL, Exeter, UK\\
$^{6}$Department of Space Studies, Southwest Research Institute, 1050 Walnut
Street, Suite 400, Boulder, CO 80302\\
$^{7}$Nordic Optical Telescope, Apartado de Correos 474, E-387~00 Santa Cruz
de la Palma, Canary Islands, Spain\\
$^{8}$Aryabhatta Research Institute of Observational Sciences, Manora
Peak, Nainital-263129, India\\
}
\begin{document}

\pagerange{\pageref{firstpage}--\pageref{lastpage}} \pubyear{2009}

\maketitle

\label{firstpage}

\begin{abstract}
We report a transit timing study of the transiting exoplanetary system
HD~189733. In total we observed ten transits in 2006 and 2008 with the 
2.6-m Nordic Optical Telescope, and two transits in 2007 with the
4.2-m William Herschel Telescope. We used Markov-Chain Monte Carlo 
simulations to derive the system parameters and their uncertainties, 
and our results are in a good agreement with previously published values. 
We performed two independent analyses of transit timing residuals to place
upper mass limits on putative perturbing planets. The results show no 
evidence for the presence of planets down to 1 Earth mass near the 1:2 
and 2:1 resonance orbits, and planets down to 2.2 Earth masses near the 
3:5 and 5:3 resonance orbits with HD~189733b. These are the strongest 
limits to date on the presence of other planets in this system.
\end{abstract}


\section{Introduction}

Ground-based radial velocity and photometric transit surveys have proved to
be the most successful methods for discovering exoplanets
over the past decade, yielding more than 400 extrasolar planets
discovered to date\footnote{The Extrasolar Planets Encyclopedia:
http://exoplanet.eu}. Most of the exoplanets detected are of Jupiter mass, 
but Earth-mass planets remain to be found. An additional planet in a 
transiting system will perturb the motion of the transiting planet,
and the interval between the mid-eclipses will not be constant. Deviations from the predicted
mid-transit times can therefore reveal the presence of other bodies in the system, or 
place limits on their existence. Short-term variations can uncover the
existence of other planets \citep{Agol05, Holman05}, moons
\citep{Sartoretti99, Kipping09} and also Trojans \citep{Ford07}, 
whereas long-term variations result from orbital decay \citep{Rasio96} and 
from orbital precession induced by another planet, stellar oblateness and general 
relativistic effects \citep{Miralda02, Heyl07}. Discovery of additional bodies
can constrain theories of planetary system formation and evolution.
In this paper we describe a transit timing study of the transiting exoplanet 
system HD~189733.

\begin{table*}
  \caption{Observations of the HD~189733 system. The UT date
is the date of the beginning of each night. The cycle number is in periods from 
the ephemeris given by \citet{Agol09}. For some nights the exposure time was changed during the
observations; this is indicated by the second value in parentheses.
The data rms is per exposure for the ratio of intensities of the target and
the comparison star. The barycentric mid-transit times of the HD~189733 system
are given with uncertainties defined as 68 per cent confidence limits.}
\label{times}
  \begin{tabular}{cllccccrc}\hline
{\bf Telescope} & {\bf UT date} & {\bf Cycle no.} & {\bf CCD window size} & {\bf
Exposure} & {\bf Data rms} & {\bf Mid-transit time}
& {\bf \hbox{$O\!-\!C$}} & {\bf Comment}\\
&&&{\bf (pixels)}&{\bf (s)}&{\bf (mmag)}& {\bf (BJD - 2450000)}&{\bf (s)}&\\
\hline 
NOT&2006 July 18     &-155&[1040:200]&2.5      &2.9&$3935.55805\pm 0.00028$&$ 38 \pm 25 $&\\
NOT&2006 August 07   &-146&[1040:200]&2.5      &2.6&$3955.52509\pm 0.00014$&$ 26 \pm 12 $&\\
NOT&2006 August 27   &-137&[1040:200]&2.5 (3.0)&2.7&$3975.49194\pm 0.00021$&$ -2 \pm 18 $&\\
WHT&2007 August 17   &+23& [1071,546]&10.0     &4.6&$4330.46305\pm 0.00042$&$-79 \pm 36 $&\\
WHT&2007 September 17&+37& [1071,546]&3.0 (3.5)&4.4&$4361.52352\pm 0.00044$&$-43 \pm 38 $&\\
NOT&2008 June 07     &+156&[1040:200]&3.5 (3.0)&2.6&$4625.53404\pm 0.00038$&$-35 \pm 33 $&partial\\
NOT&2008 June 18     &+161&[1040:200]&3.5      &2.3&$4636.62768\pm 0.00018$&$ 31 \pm 15 $&\\
NOT&2008 July 08     &+170&[1040:200]&3.5 (4.0)&2.4&$4656.59451\pm 0.00012$&$  1 \pm 11 $&\\
NOT&2008 July 17     &+174&[1040:200]&3.5 (3.0)&3.4&$4665.46951\pm 0.00029$&$ 62 \pm 25 $&\\
NOT&2008 July 28     &+179&[1040:200]&3.0      &2.3&$4676.56188\pm 0.00019$&$ 18 \pm 16 $&\\
NOT&2008 August 26   &+192&[1655:200]&3.5      &2.7&$4705.40332\pm 0.00019$&$ 15 \pm 16 $&\\
NOT&2008 September 15&+201&[1655:200]&2.5      &2.9&$4725.37064\pm 0.00053$&$ 28 \pm 46 $&partial\\ 
\hline
  \end{tabular}
\end{table*}

The HD~189733 transiting system is one of the best studied systems from
the ground. HD~189733 is a bright star with magnitude {\it V}=7.67 which is
orbited by a transiting Jupiter-mass planet in a period of
$\sim 2.22\jed{days}$ \citep{Bouchy05}, and which also has a distant
mid-M dwarf binary companion \citep{Bakos06a}. In 2006 HD~189733 was observed
with the {\it MOST} (Microvariability and Oscillations of STars) satellite and
these data were used to search for the existence of other bodies in the system. 
First, \citet{Croll07} searched for transits from exoplanets other than the known
hot Jupiter, with the result that any additional close-in exoplanets on
orbital planes near that of HD~189733b with sizes ranging from 
about 1.7 -- 3.5$\jed{R_{\oplus}}$, where$\jed{R_{\oplus}}$ is the Earth radius, 
are ruled out. Second,
an analysis of transit timing variations (TTVs) in these data has been carried out by
\citet{Miller-Ricci08} who found that there are no TTVs greater than
$\pm 45\jed{s}$, which rules out planets of masses larger than 1 and 
4$\jed{M_{\oplus}}$, where$\jed{M_{\oplus}}$ is the Earth mass, in the 2:3
and 1:2 inner resonances, respectively, and planets greater than $20\jed{M_{\oplus}}$ 
in the outer 2:1 resonance of the known planet and greater than
$8\jed{M_{\oplus}}$ in the 3:2 resonance.  

Analyses of transit times similar to \citet{Miller-Ricci08} have been
carried out for other transiting planetary systems. \citet{Steffen05} 
found no evidence for a second planet in the TrES-1 system, excluding planets down 
to Earth mass near the low-order, mean-motion resonances of  
the transiting planet. Similarly, \citet{Gibson09a, Gibson09b} 
found no evidence for additional planets down to sub-Earth masses in the interior 
and exterior 2:1 resonances of the TrES-3 and HAT-P-3 systems.

To measure  
times of mid-transits with sufficient accuracy to detect terrestrial mass
planets requires high quality photometry, free of systematic effects.
HD~189733 is known to have surface spots; 
\citet{Pont07} observed two spot events
in {\it HST} (Hubble Space Telescope) data when the flux during the transit 
changed by 1 and $0.4\jed{mmag}$. The presence of surface spots
on HD~189733 complicates any transit timing analysis \citep{Miller-Ricci08}.
The light curve can be distorted if 
the planet transits in front of a spot or due to intrinsic variability of the star.
The system parameters and the mid-eclipse times derived can then be affected 
by an inappropriate fitting model. 

Instrumental effects during transit ingress or egress can also influence the
accuracy and determination of transit times. For example,
if the transit light curve is not properly normalised 
so that all data points in egress have a flux level that is slightly 
too high, the transit time will be determined too early.
Correct normalisation is especially problematic for partial transit
light curves. Both instrumental effects and stellar variability can cause
that a light curve is improperly normalised. 

It is also important to have a light curve that is well-sampled during both ingress and
egress, because the transit timing information is contained in these parts.
When using large telescopes for such a bright star, only short exposure times are needed 
to get sufficient signal-to-noise and to avoid saturation, and so the cadence of observation is 
higher. For a given data accuracy, higher cadence leads to more accurately determined 
transit times.

In section \S\ref{Observations} we describe our observations, and in section
\S\ref{Data_reduction} we present our data reduction.
In section \S\ref{model} we explain the techniques used to estimate
uncertainties in our data and to measure the system parameters. Finally, in section
\S\ref{results} we describe the 3-body simulations used to place limits on the 
existence of other bodies in the system, and we conclude and discuss our
results in section \S\ref{conclusions}.

\section{Observations}\label{Observations}

We observed eight full and two partial transits of HD~189733 with the 2.6-m Nordic 
Optical Telescope (NOT), La Palma, Spain, using ALFOSC (the Andalucia Faint 
Object Spectrograph and Camera), and two full transits using the AG2 camera 
on the 4.2-m William Herschel Telescope (WHT) of the Isaac Newton Group (ING), 
La Palma, Spain (Table~\ref{times}).

ALFOSC has a $2048\times 2048$ back-illuminated CCD with scale 
$0.19\jed{arcsec/pixel}$ and field of view (FOV) $6.5\times 6.5\jed{arcmin^2}$. 
To reduce the readout time of each exposure and the duty cycle of
observation we windowed the CCD with the window sizes summarized in Table~\ref{times}.
We used a Str\"{o}mgren y filter to minimize effects of colour-dependent atmospheric 
extinction on the differential photometry and the effect of limb darkening on the
transit light curves. We defocused the telescope typically to $3.4\jed{arcsec}$, spreading 
the light inside full width at half maximum (FWHM) of the 
Point Spread Function (PSF) over $\sim$ 250 pixels, 
in order to minimize the impact of pixel-to-pixel sensitivity variations, 
and to prevent saturation. Exposure times were chosen to keep counts below
50,000 
per pixel to avoid saturation of features such as hot spots and speckles in the defocused
stellar images, and to ensure data linearity. The typical exposure time for
the NOT data was $3\jed{s}$ (Table~\ref{times}). 

AG2 is a frame-transfer CCD mounted at the WHT's folded Cassegrain focus, 
based on an ING-designed 
autoguider head. The FOV is $3.3\times 3.3\jed{arcmin^2}$ and the
scale is $0.4\jed{arcsec/pixel}$. We used a Kitt Peak R filter and
defocused the telescope to 10 and $12\jed{arcsec}$ for the two nights,
spreading the FWHM-light over $\sim$ 490 and 700 pixels, respectively.
The corresponding exposure times were 10 and $3\jed{s}$.

The mid-time of each exposure was  
converted to the Barycentric Julian Date (BJD) using the program 
BARCOR\footnote{http://sirrah.troja.mff.cuni.cz/\~{}mary}.
We use BJD throughout this paper, because for this system the
Heliocentric Julian Date would accumulate an error of up to 4 seconds.

\section{Data Reduction}\label{Data_reduction}

Bias subtraction, flat-field correction and aperture photometry was
performed using standard IRAF\footnote{The Image Reduction and Analysis Facility
(IRAF) is distributed by the National Optical Astronomy Observatories, which
are operated by the Association of Universities for Research in Astronomy,
Inc., under cooperative agreement with the National Science Foundation.}
procedures.

To ensure a signal-to-noise in excess of 1,000 in our Str\"{o}mgren y-filter flat fields
for the NOT data we generated a master flat field for each night using individually weighted
normalised flat fields from the entire observing season combined with weights 
$W=1-D/S$, where $D$ is the time interval between each night and 
date of observation, and $S$ is the season length. Applying 
flat-field corrections has only a minor effect on the resulting NOT
photometry, because of the heavily defocused PSF.

For the WHT data we determined master flat fields with a signal-to-noise
greater than 1,000 for both nights. However, we identified a position-angle dependent 
scattered light component in the flat fields, which introduced systematic noise in our  
WHT photometry. Therefore we did not apply flat-field corrections.

We used the star 2MASS 20003818+2242065 as our comparison star for the WHT
data. In our NOT data there are two available comparison stars, 2MASS~20003818+2242065
and 2MASS~20003286+2241118. We found the ratio of their measured intensities varies
by a few mmag with time, as the telescope tracks across the meridian.
This variation correlates with small drifts in the positions of the
stars on the CCD, suggesting that some light is being lost from the aperture around
one of the stars due to the wings of the PSF 
drifting out of that aperture. A similar variation is seen for the ratio 
of the intensities of 2MASS~20003286+2241118 and out-of-transit HD~189733, 
but not for 2MASS~20003818+2242065 and HD~189733, suggesting that it is light from
2MASS~20003286+2241118 which is being lost. This star is the farther of the two from 
HD~189733, and we conclude that the variation in measured intensity is due to
a combination of the small drifts in stellar position on the CCD,
and the variation of the defocused PSF across the FOV.
We therefore used only the comparison star which is closer to HD~189733.

We used circular, equal diameter, photometric apertures  
for both HD~189733 and the comparison star. A range of aperture sizes was
tried and that producing the minimum noise in the out-of-transit data was adopted
and fixed during each night. 
The aperture radius for all stars ranged from 18 -- 29 pixels for different 
NOT nights and the typical FWHM was around 18 pixels ($3.4\jed{arcsec}$).
For the two WHT nights the aperture radius was 28 and 30 pixels,
respectively, and the corresponding FWHM was 25 and 30 pixels
(10 and $12\jed{arcsec}$). 

We ensured the apertures tracked small drifts in the stellar positions on
each image by using a large centroiding box of size $4\,\times$~FWHM. 
During each night drifts in the stellar positions on the CCD
were less than 7 (NOT) and 4 pixels (WHT).
The sky background was subtracted using an estimate of its brightness determined within 
an annulus centred on each star with a width of 10 pixels. For each night, 
differential photometry was computed by taking the ratio of counts from
HD~189733 to the counts from the comparison star. 
We normalised our data using linear fits that were
computed together with other system parameters as described in \S\ref{model}.
  
\begin{figure*}
 \resizebox{17cm}{!}{\includegraphics{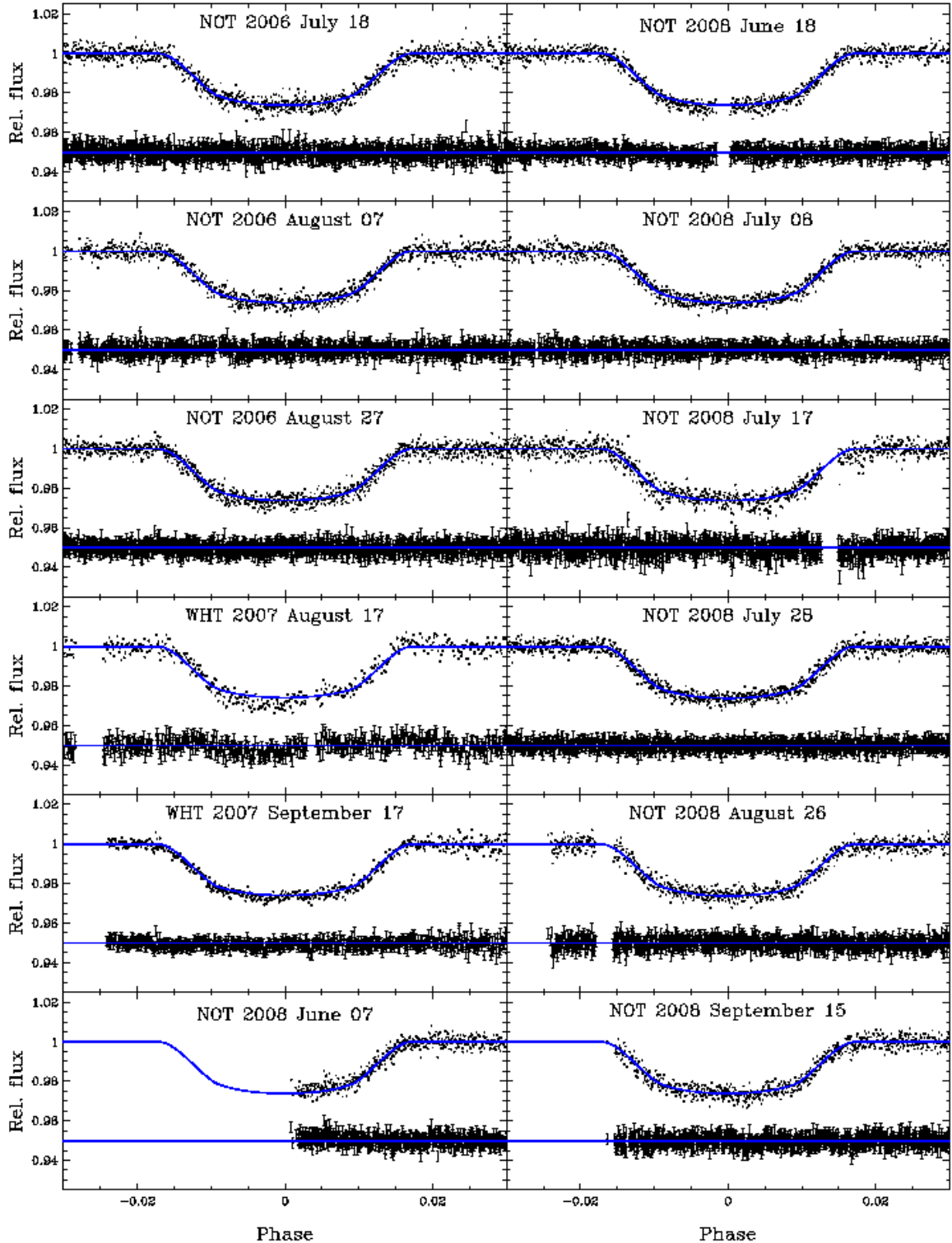}}
\caption[ ]{Differential photometry of the HD~189733 system over-plotted with the 
best-fitting model (solid line) from the MCMC fit. The residuals and $1\sigma$ 
error bars are also plotted, offset by a constant flux for clarity. The phase was
computed using best-fitting transit times presented in Table~\ref{times}.
The photometry for NOT data is unbinned and for WHT is binned in time
with 10 second bins to give the similar cadence as for NOT data for clarity.}
\label{plot1}
\end{figure*}

The normalised unbinned NOT light curves and binned WHT light curves,
averaged into 10-second bins to have the similar cadence as the NOT data, are
shown in Fig.~\ref{plot1} along with their best-fitting models,
residuals and data error bars, as derived in \S\ref{model}.

\section{Light-curve modelling}\label{model}

To estimate the system parameters we used a parametrized model where we assumed 
a circular orbit around the centre of mass to calculate the normalised separation, 
$z$, of the planet and star centres as a function of time. The analytic formulas 
of \citet{Mandel02} were used to calculate the fraction of the stellar flux 
occulted by the planet using $z$ and the planet-to-star radius ratio, $\rho$. 
We assumed a quadratic limb darkening law:
\begin{equation}
\frac{I_{\mu}}{I_0}=1-u_1 (1-\mu)-u_2 (1-\mu)^2,
\end{equation}
where $I$ is the intensity, $\mu$ is the cosine of the angle between the
line of sight and the normal to the stellar surface, and $u_1$ and $u_2$ are the
linear and quadratic limb darkening coefficients. For the NOT data we allowed the limb
darkening coefficients to be free parameters, in order to include possible
errors in the limb darkening coefficients into our final system parameters
and mid-transit times. For the WHT data we adopted values $u_1=0.4970$ and
$u_2=0.2195$ from the tables of \citet{Claret00} and fixed them in the subsequent 
analysis. These correspond to the Johnson R filter which has similar characteristics 
as the Kitt Peak R filter used. 

To compute our model we folded all the NOT light curves of full transit except
for the night 2008 July 17 which displays obvious systematic changes during transit.
In our photometry we cannot easily distinguish spot effects from systematic 
instrumental errors; to do so would require the instrumental 
systematic noise to be much less than the predicted spot signatures.  
We fitted simultaneously planetary and stellar radius, 
$R_p$ and $R_{\star}$, respectively, the orbital inclination, $i$, two limb
darkening coefficients, $u_1$ and $u_2$, transit time, $T_{0,n}$, 
and additional two parameters for each night $n$ -- the out-of-transit flux,
$f_{oot,n}$, and a time gradient, $t_{Grad,n}$. These two parameters were allowed
to be free to account for any normalisation errors in the data.
For each change of $R_{\star}$, the stellar mass, $M_{\star}$, was recomputed
using the scaling relation $R_{\star}\propto M_{\star}^{1/3}$. We fixed the planetary
mass value $M_p=1.15\pm 0.04\jed{M_J}$ \citep{Bouchy05}, adopted a  
period $P=2.21857503\pm 0.00000037\jed{d}$ \citep{Agol09}, and using
Kepler's third law we updated the orbital semi-major axis for each choice of
$M_{\star}$.

We ran Markov-Chain Monte Carlo (MCMC) simulations \citep{Tegmark04,
Ford06, Holman06} with the Metropolis--Hastings algorithm \citep{Ford05} 
to estimate the best-fitting parameters and their uncertainties. From an
initial point, a chain is generated by iterating a jump
function, which adds a random value selected from a Gaussian
distribution with mean 0 and a standard deviation 1, scaled by a factor
specific for each parameter so that $\sim 44$ per cent of each parameter sets are
accepted \citep{Gelman03, Ford06}. In each step of the generated chain 
the $\chi^2$ fitting statistic for old and new parameter values is computed:
\begin{equation} \label{chi}
\chi^2=\sum\limits _{i=1}^{N_{DOF}}
\left[\frac{f_{i}(obs)-f_{i}(theor)}{\sigma_i}\right]^2+\frac{(M_{\star}-M_0)^2}{\sigma_{M_0}^2}.
\end{equation}
Here $f_{i}(obs)$ is the flux observed at time $i$, $\sigma_i$ is the 
corresponding uncertainty, $f_{i}(theor)$ is the flux calculated using
formulas of \citet{Mandel02} and $N_{DOF}$ is the number of
measurements in each light curve. The new parameter is then accepted if its $\chi^2$
is lower than that for the previous parameter, or accepted with a probability
$p=\exp\left(-\Delta\chi^2/2\right)$ if its $\chi^2$ is higher.
The second term in Eq.~(\ref{chi}) is a Gaussian prior placed on
$M_{\star}$, where $M_0=0.82\jed{M_{\odot}}$ and
$\sigma_{M_0}=0.03\jed{M_{\odot}}$ is the stellar mass and its uncertainty, 
estimated from stellar spectra by \citet{Bouchy05}. This ensures that errors 
in the stellar mass, which are the greatest source of uncertainty when
deriving the system parameters and transit times, are taken into account.

The scale factors were
chosen so that $\sim 44$ per cent of parameter sets were accepted \citep{Gelman03,  
Ford06}. For each simulation we created 10 independent chains, with length
at least 100,000 points per chain to ensure convergence. Each chain was
initiated by a parameter that was within $\pm 5\sigma$ of a previously known 
best-fitting parameter value using the estimated uncertainty $\sigma$. The first 20 per cent 
of each chain was discarded to minimize the effect of the initial conditions. 
We checked convergence of generated chains using the \citet{Gelman92} R statistic 
and created chains until $R<1.03$, a good sign of convergence.  

To estimate appropriate error bars in our data accounting for any correlated noise, 
we used a procedure similar to that of \citet{Gillon06} and \citet{Narita07}. We assigned the same
error bars to all the data points including only Poisson noise. An initial MCMC analysis 
of the folded NOT light curves was used to estimate the parameters $R_p$,
$R_{\star}$, $i$, $u_1$, $u_2$, $T_{0,n}$, $f_{oot,n}$ and $t_{Grad,n}$. The first model
light curve was used to find the differences between the data and the model for each 
individual night. Then we rescaled the error bars to satisfy the condition 
$\chi^2/N_{DOF}=1.0$, where $N_{DOF}$ is the number of measurements in each light curve. 
For the night of 2008 July 17, the nights of the two partial transits 
(2008 June 07 and 2008 September 15) and for the WHT light curves (2007
August 17 and 2007 September 17) we adopted our first model
and ran MCMC analysis to find initial parameters $T_0$, $f_{oot}$ and $t_{Grad}$
for each night independently. Then we rescaled the error bars similarly as
before. We assume that our initial model is a good description of the light
curve. Compared to this model, we found that for the NOT data errors are higher 
by factors of 2.3 -- 3.4 than errors including only Poisson noise, and for the WHT data by 
factors of 4.6 and 4.4 for the two nights, respectively. The data rms errors per
exposure are presented in Table~\ref{times}. The predicted 
rms due to photon noise, which is dominated by the fainter comparison
star, and to atmospheric scintillation, is $\sim 2.5\jed{mmag}$ for the NOT data
and $\sim 3\jed{mmag}$ for the WHT data.

The amplitude of systematic trends in the photometry was estimated from the
standard deviation over one residual point, $\sigma_1$, and from the
standard deviation of the average of the residuals over $N$ successive
points, $\sigma_N$. We solved the following system of two equations
given by \citet{Gillon06}:
\begin{equation}
\sigma_1^2=\sigma_w^2+\sigma_r^2,
\end{equation}
\begin{equation}
\sigma_N^2=\frac{\sigma_w^2}{N}+\sigma_r^2,
\end{equation}
to obtain the amplitude of the white noise, $\sigma_w$, which is
uncorrelated and averages down as $(1/N)^{1/2}$, and the red noise,
$\sigma_r$, which is correlated and remains constant for specified $N$. 
The error bars were then adjusted by multiplying by $[1+N(\sigma_r/\sigma_w)^2]^{1/2}$
and these rescaled uncertainties were used for the subsequent fitting
procedure. To account properly for the systematic errors, the resulting multiplying
factor was computed as the average of values using different $N$ in the range 15 --
30 minutes (the typical time-scale of ingress and egress).

To create our final model, we proceeded as before but this time including
systematic noise in our data and therefore properly estimating parameter
uncertainties. We ran MCMC using the folded NOT light curves and fitting the
parameters as described earlier. We created 10 chains, each with length
2,000,000 points in order to achieve convergence.
Ultimately, we used our final model to find 
individual mid-eclipse times and two normalisation parameters 
for the night of 2008 July 17, the nights of the two partial transits 
(2008 June 07 and 2008 September 15) and for the WHT light curves (2007
August 17 and 2007 September 17).

\section{Results}\label{results}

\begin{figure*}
 \resizebox{16.3cm}{!}{\includegraphics{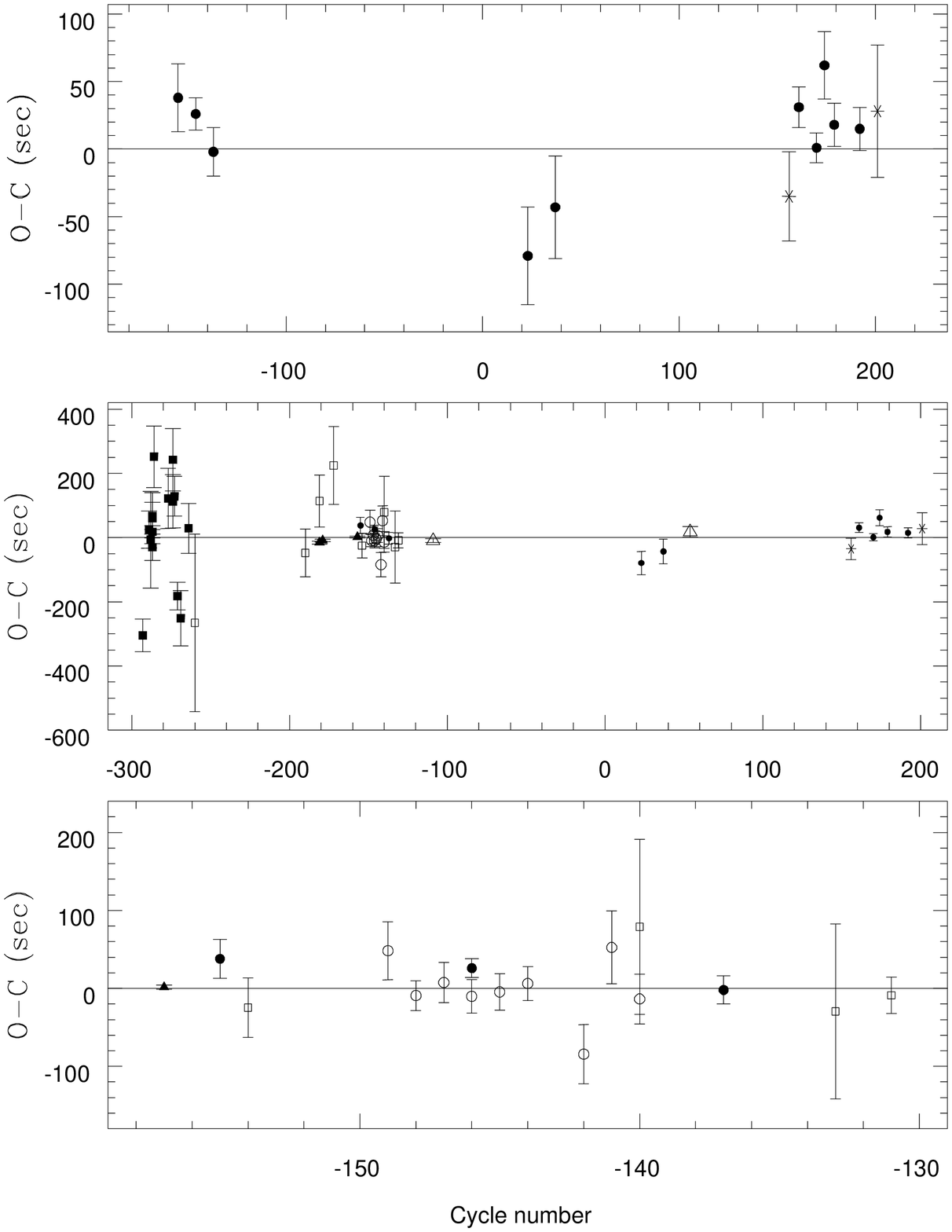}}
\caption[ ]{{\it Top:} NOT and WHT \hbox{$O\!-\!C$} residuals of mid-transit times 
of the HD~189733 system including both partial (star symbol) and full transits 
(filled circle symbol). 
{\it Middle:} Previously published values plotted together with NOT and WHT
results. Filled squares: \citet{Bakos06b}, ground-based; filled triangles:
\citet{Pont07}, {\it HST}; open squares: \citet{Winn07}, ground-based;  open circles:
\citet{Miller-Ricci08}, {\it MOST}; open triangles: \citet{Knutson09}, {\it Spitzer}; 
filled circles and stars: this work. 
{\it Bottom:} The same as the middle but zoomed for clarity.
The cycle number is in periods from the ephemeris given by \citet{Agol09}.
A horizontal line is plotted in each panel at \hbox{$O\!-\!C =0$} to guide
the eye. Our timing measurements are the most accurate from known ground-based 
observations.}\label{oc}
\end{figure*}

The final system parameters are presented in Table~\ref{param} and are consistent within 
$\sim 2\sigma$ error bars with the previously published values
\citep{Bakos06b, Pont07, Winn07,
Miller-Ricci08}. The resulting limb darkening coefficients for the NOT data
were $u_1=0.46 \pm 0.10$ and $u_2=0.35 \pm 0.13$. 

The final barycentric transit times can be found in Table~\ref{times}.
The uncertainties are defined as 68 per cent confidence limits. To compute the
observed-minus-calculated values (\hbox{$O\!-\!C$}) we used the ephemeris
given by \citet{Agol09}:
\begin{equation}
T_c(E)={\rm HJD}\,(2454279.436741\pm 0.000023)+
\end{equation}
$(2^{\rm d}\!\!.21857503\pm 0^{\rm d}\!\!.00000037)\times
E.\label{ephemeris}$\\
The resulting \hbox{$O\!-\!C$} residuals together with all the other
previously published values \citep{Bakos06b, Pont07, Winn07, Miller-Ricci08, Knutson09} 
are plotted in Fig.~\ref{oc}. Our observations did not bring any refinement of the ephemeris and we 
confirm that presented by \citet{Agol09}.
For the night 2006 August 07 a transit timing
measurement of HD~189733 was also presented by \citet{Miller-Ricci08} from the {\it MOST}
data and it is consistent within $2\sigma$ error bars with our measurement.

\begin{table}
  \begin{center}   
  \caption{System parameters of HD~189733. The uncertainties are 
68 per cent confidence limits.} \label{param}
\begin{tabular}{lccc}\hline
{\bf Parameter} & {\bf Symbol} & {\bf Value} & {\bf Units}\\
\hline
Planet radius&$R_p$&$1.142\pm 0.014$&$\jed{R_J}$\\
Star radius&$R_{\star}$&$0.755\pm 0.009$&$\jed{R_{\odot}}$\\
Orbital inclination&$i$&$85.70\pm 0.03$&deg\\ 
Planet/star radius ratio&$\rho$&$0.1556\pm 0.0027$&\\
Total transit duration&$T_d$&$1.807\pm 0.023$&h\\
Impact parameter&$b$&$0.667\pm 0.009$&\\
\hline
  \end{tabular}
 \end{center}
\end{table} 


\subsection{Transit timing variations analysis}

For all our observations which span more than two years, the mean \hbox{$O\!-\!C$}$=5\pm 38\jed{s}$, where
the quoted error is the rms scatter in the \hbox{$O\!-\!C$} values and is
slightly larger than the average \hbox{$O\!-\!C$} uncertainty $\sim 25\jed{s}$. 
None of our \hbox{$O\!-\!C$} measurements is a significant outlier. The two
largest \hbox{$O\!-\!C$} values for the nights 2007 August 17 and 2008 July 17 
coincide with obvious systematic changes during the transit (see Fig.~\ref{plot1})
and both have the same or larger uncertainty than the average value. Therefore the
rms scatter in the \hbox{$O\!-\!C$} values of 38 s is a good estimate for
placing limits on the presence of other planets in the system.   

We used this conclusion to place mass limits on the existence of planets on orbits interior and exterior to 
HD~189733b. First, we selected the mass, semimajor axis and eccentricity of the
putative perturbing planet. The orbital inclination was set so that HD~189733b and
the perturbing planet have coplanar orbits. The two-planet system was then
numerically integrated using the Bulirsch-Stoer integrator \citep{Press92}.
We determined all mid-transit times of HD~189733b over a time-span of 500 days, 
an interval long enough to cover at least 14 orbits of all perturbing
planets we can exclude, and used these data to estimate TTVs. 
The mass, initial semimajor axis and initial eccentricity of the perturbing planet 
were varied to determine the TTV amplitude for different planetary
configurations.

\begin{figure*}
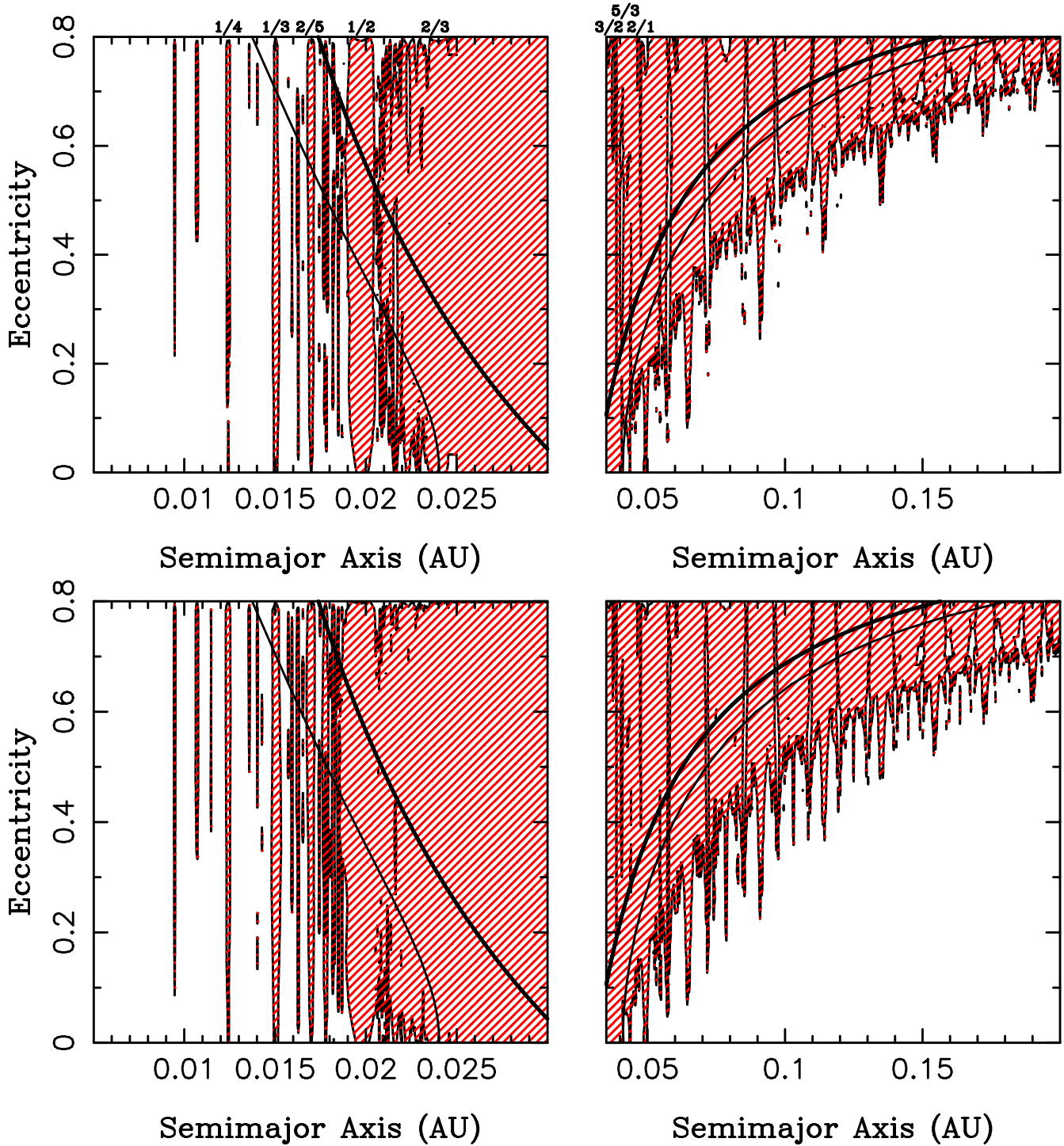

 \resizebox{\hsize}{!}{\includegraphics{fig3.ps}}
 \hfill
 \resizebox{\hsize}{!}{\includegraphics{fig4.eps}}
\caption[ ]{A numerical survey of the HD~189733 system showing 38-s TTVs
caused by an inner (left plots) and outer (right plots) 1 Earth-mass planet
($m_2=3\times10^{-6}\jed{M_{\odot}}$, top) and 2 Earth-masses planet
($m_2=6\times10^{-6}\jed{M_{\odot}}$, bottom). 
The shaded area excludes a range of possible eccentricities and semimajor
axes for a putative 1 and 2 Earth-masses inner/outer planet in the system based on our 
observational non-detection of TTVs greater than $\pm 38\jed{s}$. We do not display 
plots for Jupiter-mass planets as these would easily be
detected in radial velocity searches. The thick solid line shows a
boundary where a collision between the two planets can occur. It is
defined so that the apocentre/pericentre of the inner/outer perturbing planet
equals to the semi-major axis of the transiting planet.
The thin solid line represents a Hill stability computed according to
\citet{Gladman93}. On the top of the upper panel we indicate the major
resonances of the putative perturbing planet and HD~189733b.}\label{ttv}
\end{figure*}

Fig.~\ref{ttv} shows the range of the inner and outer planet's orbits that produce
TTVs smaller/larger than $\pm 38\jed{s}$ and are thus  
compatible/incompatible with our TTV observations of the HD~189733 system.
The shaded area in Fig.~\ref{ttv} excludes a range of possible eccentricities 
and semimajor axes for a putative 1 Earth-mass (top) and 2 Earth-masses
(bottom) inner (left) and outer (right) planet in the system. Based on this
analysis, our observations of the HD~189733 system show no evidence for the presence 
of planets down to 1 Earth mass in the 2:1, 3:2 and 5:3
exterior resonance orbits, planets down to 1 Earth mass in the 1:2, 1:3,    
2:3 and 2:5 interior resonance orbits, and planets down to 2 Earth masses
in the 1:4 interior resonance orbit with HD~189733b.
However, not all of these resonant orbits are Hill stable. We computed Hill
stability according to Eq.~(21) of \citet{Gladman93} for both inner and
outer perturbing planet and displayed the result in Fig.~\ref{ttv} using thin solid line.
For the inner/outer perturbing planet all the orbits on the left/right to the thin
solid line are Hill-stable, which means that close approaches between two
planets are forbidden. For the rest of the parameter space the Hill 
stability of the system is unknown; the system still may be Hill-stable.    

\citet{Nesvorny09} showed that the TTV signal can be significantly amplified for 
planetary systems with substantial orbital inclinations of the transiting
and perturbing planet and/or in the case of transiting planet in an
eccentric orbit with an anti-aligned orbit of the perturbing planetary companion. 
Therefore for most orbital architectures of exoplanetary 
systems we determine the perturber's upper mass from our TTVs under the assumption 
of coplanar orbits of transiting and perturbing planets. 

However, the above mentioned analysis does not take into account time
sampling of our measured transit times and their uncertainties. It is possible 
to have a system whose TTV amplitude exceeds 38 s but remains consistent with the 
available transit timing data. To assure that the limits on additional
planets presented in this paper are not overestimated, in addition to our
previous analysis we compared model timing residuals
against the transit times to place upper mass limits for a putative
perturbing planet. We used the same procedure as \citet{Gibson09a, Gibson09b},
where more details can be found. To compute model timing
residuals we integrated the equations of motion for a three body system
using a 4th-order Runge-Kutta method, with the first two bodies representing
the star and planet of the HD~189733 system, and the third body representing
a putative perturbing planet. The transit times were extracted when the star
and transiting planet were aligned along the direction of observation, and
the residuals from a linear fit were taken to be the model timing residuals.
For each model, TTVs were extracted for 6 equally spaced directions of
observations, and we simulated 3 years of TTVs to cover the full range of observations.
The resulting TTVs were then compared to transit times presented in the middle panel 
of Fig.~\ref{oc}, i.~e. all available transit times. Due to computational limitations we assumed 
that the amplitude of the timing residuals scales proportionally to the perturber 
mass \citep{Agol05, Holman05}, that a perturber has circular orbit and that the 
orbits of the planets are coplanar.

We created models with period ratios in the
range 0.2 -- 5.0, increasing the sampling around the interior 1:2 and
exterior 2:1 resonances. The maximum allowed mass for each model was
calculated as in \citet{Gibson09a, Gibson09b}. We scaled the perturber mass
until the $\chi^2$ of the model fit increased by a value $\Delta \chi^2 = 9$
\citep{Steffen05} from that of a linear ephemeris. Then we minimized the $\chi^2$
along the epoch, and rescaled the perturber mass again until the maximum allowed mass 
was determined. This was repeated for each direction of the observation, and the
maximum mass found was set as our upper mass limit. This process was repeated twice
to verify our assumption that the timing residuals scale proportionally 
with the mass of the perturbing planet.

\begin{figure*}
 \resizebox{14cm}{!}{\includegraphics{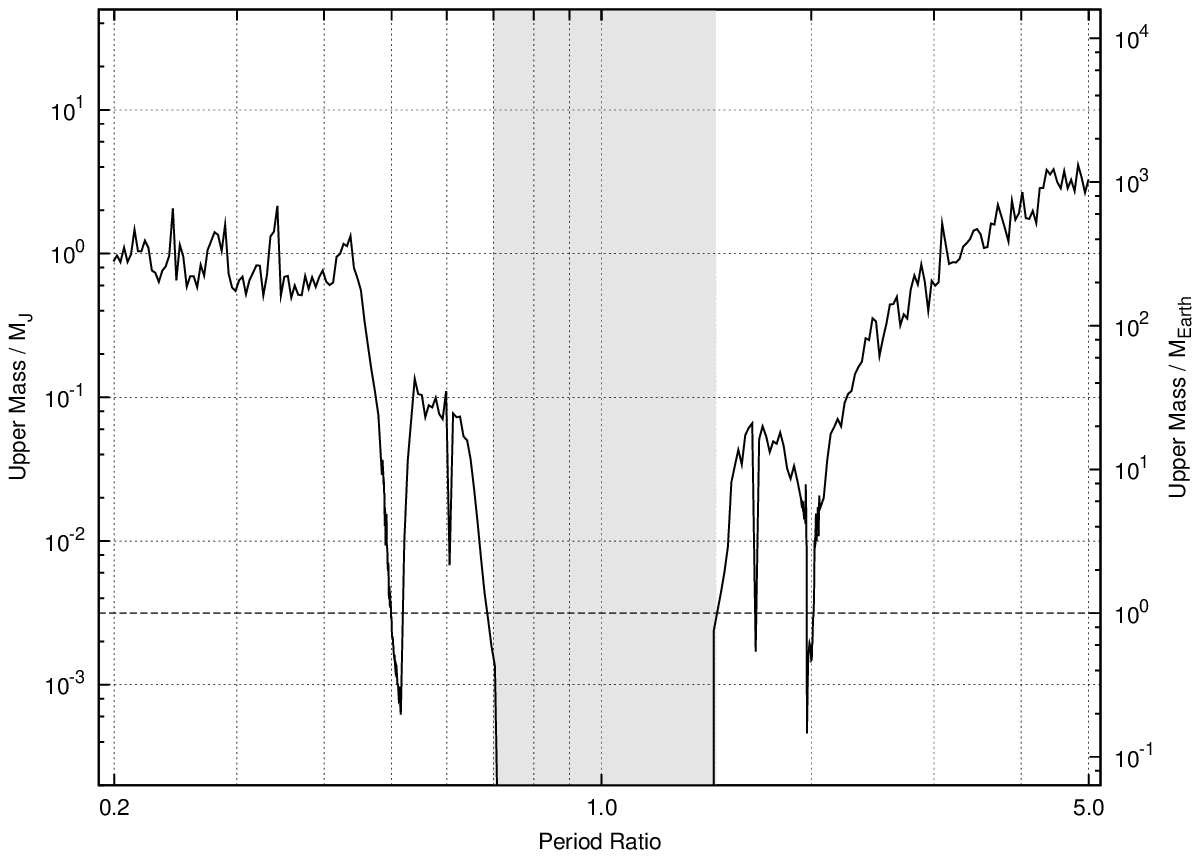}}
\caption[ ]{Upper mass limits on a putative second planet in the HD~189733
system as a function of period ratio based on the comparison of model timing
residuals and all available transit times. The solid line represents the upper mass found
using three-body simulations. The horizontal dashed line shows an Earth-mass
planet. The grey area is the region where an Earth-mass perturbing planet is not guaranteed 
to be Hill stable.}\label{ttv2}    
\end{figure*}

The resulting
upper mass limits are plotted as a function of the period ratio in Fig.~\ref{ttv2}.
The solid line represents the upper mass limits from our three-body
simulations, and the horizontal dashed line shows an Earth-mass planet.
Based on this analysis, the available data were sufficiently sensitive to probe for
masses as small as 0.2 and $0.15\jed{M_{\oplus}}$ near the interior 1:2 and
exterior 2:1 resonances with HD~189733b, respectively. The corresponding upper 
masses near the 3:5 and 5:3 resonances with HD~189733b are 2.2 and 
$0.54\jed{M_{\oplus}}$. In the rest of the space outside the region between
the 2:3 and 3:2 resonances with HD~189733b the upper
mass limits are of the order of a few tens of Earth masses to a few Jupiter
masses. However, these upper mass limits result from the assumption of a
circular orbit of a perturber. Eccentric orbits may lead to smaller TTVs,
and hence planets larger than our upper mass limits in eccentric orbits
could exist in these regions. Unfortunately, accounting for eccentric orbits
is computationally unfeasible using these models due to a large parameter
space. Thus real upper mass limits of a perturber in a low eccentric
orbit can be as much as an order of magnitude larger \citep{Gibson09b}.

We also consider the possible presence of Trojans
in the system. According to \citet{Ford07} transit times are the same for
a system without a Trojan and for a system where the transiting planet and Trojan 
have equal eccentricities and the Trojan resides exactly at the Lagrange L4/L5 fixed
point. TTV analysis alone is not suitable for constraining the presence of
Trojans in transiting systems. However, a
comparison of the photometrically observed transit time and the transit time calculated 
from the radial velocity data assuming zero Trojan mass can reveal a Trojan 
or place upper limits on its mass. Such an analysis was done by \citet{Madhusudhan09} 
who found an upper limit of $22\jed{M_{\oplus}}$ for a Trojan in the HD~189733 system. 
In addition, \citet{Croll07} searched for Trojan transits in {\it MOST}
photometry, assuming similar inclinations of the Trojan's and transiting
planet's orbits, and concluded that Trojans with a radius above
$2.7\jed{R_{\oplus}}$ should have been detected with 95 per cent confidence.
Using a mean density of $\rho \sim 3000\jed{kg\,m^{-3}}$, this corresponds to
$11\jed{M_{\oplus}}$. We used Eq.~(1) of \citet{Ford07} to estimate what
Trojan's mass can be excluded in the system based on 38 s rms of our TTVs. However, the
amplitude of the angular displacement of a putative Trojan from the Langrange point 
is not known. If these libration amplitudes are similar as for Trojans
orbiting near the Sun-Jupiter Langrange points, that is 5 -- 30 deg \citep{Murray00}, 
our TTVs show no evidence for Trojans with masses higher than
$5.3\jed{M_{\oplus}}$.

For an Earth-mass exomoon in a circular orbit about HD~189733b \citet{Kipping09} 
predicted TTV amplitude of 1.51 s and transit duration variation (TDV) 
amplitude of 2.94 s. Increasing the eccentricity of the moon's orbit 
decreases TTV amplitude, but increases TDV amplitude. However, for the HD~189733 
system these variations are too small to be detectable in our data. 

\section{Conclusions and discussion}\label{conclusions}

\citet{Miller-Ricci08} found no TTVs greater than $\pm 45\jed{s}$ in {\it
MOST} data, and excluded super-Earths of masses larger than 1 and
4$\jed{M_{\oplus}}$ in the 2:3 and 1:2 inner resonance, respectively, 
and planets greater than $20\jed{M_{\oplus}}$ in the outer 2:1 resonance 
of the known planet and greater than $8\jed{M_{\oplus}}$ in the 3:2 resonance.
\citet{Miller-Ricci08} assumed that the orbit of the perturbing planet is
circular and that additional planets in eccentric orbits would produce stronger perturbations.
However, \citet{Nesvorny09} showed that an eccentric planet can produce 
stronger or weaker perturbations depending on the relative angular position 
of its orbital pericentre. 

In this paper we used two different methods to determine the upper mass limits 
for a putative perturbing planet in the HD~189733 system and thus the results 
of both analyses can be directly compared. Our first analysis does not take
into account time sampling of the measured transit times and their uncertainties.
On the other hand, it was possible to probe for eccentric orbits of a perturbing 
planet, which is more rigorous than assuming a circular orbit \citep{Nesvorny09}.
Further analysis was performed to assure that the limits on additional planets  
presented in this paper are not overestimated. Unfortunately, applying this
mothod for eccentric orbits of a perturbing planet would increase the number 
of parameters enormously, thus we assumed a circular orbit for the
perturber.

Due to the limitations of our TTV analyses, we adopt the least constaining 
limits to conclude what upper masses of a putative perturbing planet can be 
excluded in the HD~189733 system. The results show no evidence for the presence of
planets down to 1 Earth mass near the 1:2 and 2:1 resonance orbits, and planets
down to 2.2 Earth masses near the 3:5 and 5:3 resonance orbits with HD~189733b. 
These are the strongest limits to date on the presence of other planets in this 
system, based on results of two independent TTVs analyses. We also discuss 
the possible presence of Trojans in the system, and conclude that the highest 
limit on a Trojan mass is $5.3\jed{M_{\oplus}}$ if its libration amplitude is 
similar as for Trojans orbiting near the Sun-Jupiter Lagrange points.

\section*{Acknowledgments}

We would like to thank the anonymous
referee for useful suggestions and improvements. We are grateful 
to P.~Harmanec and R.~Karjalainen for their careful reading of the manuscript 
and comments. The data presented here have been taken using ALFOSC, which is 
owned by the Instituto de Astrofisica de Andalucia (IAA) and operated at the 
Nordic Optical Telescope under agreement between IAA and the NBIfAFG of the
Astronomical Observatory of Copenhagen.
The research was supported by the grants 205/08/H005 and 205/06/0304 
of the Czech Science Foundation and from the Research Program 
MSM0021620860 of the Ministry of Education of the Czech Republic.
We acknowledge the use of the electronic bibliography maintained 
by NASA/ADS system and by the CDS in Strasbourg.

{}

\label{lastpage}

\end{document}